\documentclass[pra,showpacs,twocolumn]{revtex4}
\usepackage{graphicx}
\usepackage{amssymb}
\usepackage{amsmath}
\usepackage{amsbsy}
\usepackage{color,soul}
\usepackage{upgreek}
\usepackage{siunitx}
\usepackage{tikz}
\usepackage{soul}

\usepackage{multirow}
\usepackage{enumitem}

\usepackage{hyperref}

\hypersetup{
    colorlinks=true,
    linkcolor=blue,
    filecolor=magenta,      
    urlcolor=blue,
    citecolor=red
}
 
\usepackage{color,soul}

\usepackage{xcolor}

\begin{document}



\title{CPU-efficient numerical code for charged particle transport through insulating  straight capillaries}

\author{Giglio Eric}


\date{\today}
\affiliation{Centre de Recherche sur les Ions, les Mat\'eriaux et la Photonique (CIMAP), Normandie Univ, ENSICAEN, UNICAEN, CEA, CNRS, F-14000 Caen, France}






\begin{abstract}
A numerical code, labeled InCa4D, used for simulating CPU-efficiently the guiding of charged beam particles through insulating straight nano or macro capillaries, is presented in detail. The paper may be regarded as a walk through the numerical code, where we  discuss  how we compute the charge deposition and charge dynamics at the interfaces of a straight  capillary and how we compute the electric field with imposed boundary conditions. The latter add surface polarization charges at the dielectric interfaces  and  free charges at conducting interfaces.
Absorbing boundary conditions allow for a leakage current.
As a result, the electric field in InCa4D yields accurate relaxation rates and decay rates for both cases, namely where the outer surface of the straight capillary is covered by a grounded conducting paint or not. Eventually, we show how  we sample the initial conditions of the inserted beam particles  and how we evaluate CPU-efficiently the particles' trajectory, allowing to compute typically $10^6$ trajectories in about two hours on a modern CPU.
\end{abstract}

\maketitle


\section{Introduction}

Since the pioneering works of Ikeda \cite{Ikeda} and Stolterfoht \cite{Stolterfoht_2002}, the guiding phenomenon  of charged particles through nano or macro capillaries has attracted much attentions \cite{Lemell,Stolter_report} and is still an active research field, as shown by recent publications \cite{Nguyen,Shidong,Giglio_image_force}. 
On the theoretical side, simulations of beam particles (electrons, positrons or ions)  through insulating capillaries provided quite early fruitful theoretical support to experimental results and are still improving the quality of their predictions.  
The numerical codes used by the authors for the simulations are usually home-made. 
The  codes are typically  segmented into two parts. One part calculates
the trajectories of the inserted beam particles through the capillary until they escape or hit the inner capillary surface, depositing some charge at the impact point.  The second part calculates the time evolution of the deposited charges in the capillary wall. Both parts are coupled by the electric field generated by the accumulated charges in the capillary.

There exist different approaches for the charge dynamics in insulating capillaries. On one hand,  the charge that accumulates in the capillary is described by a set of $N$ point charges. Depending on the model, the point charges are field driven along the surface or even into the bulk or simply sit at their impact point. The point charges then decay  in time with a decay rate  proportional to the bulk conductivity of the insulator. The electric potential at a given position is evaluated by summing over all (non-zero) point charges, 
$ V(\vec{r}) = \sum_i \frac{2}{1+ \epsilon_r} \frac{q_i}{|\vec{r_i}-\vec{r}|}$, where a screening factor accounting for surface polarization is usually added. Such approaches were for example used in \cite{Schiessl05,Schweigler_NIMB,Niko_PRA_2014,
Niko_NIMB_2015,Niko_PRA_2015,Stolterfoht13-1,Stolterfoht13-2,niko_atoms_2020}.

On the other hand, the accumulated charges are described by a surface charges represented on axis symmetric 2D grids at the inner and outer surfaces of the capillary. The charges that are deposited at the impact point are smeared over nearest grid points. The time evolution of both surface charge distributions is described by two coupled continuity equations. The electric field that drives the  surface charges and the trajectory of the inserted beam particles is deduced from the free and induced polarization surface charges at the inner and outer surface of the capillary.

For the charge dynamics, the grid approach has some advantages over the point charge approach. 
First, boundary conditions at dielectric and conducting interfaces can easily be imposed on the electric potential and electric field, which allows calculating explicitly polarization charges that appear at the vacuum-dielectric interfaces and  free charges that appear at the vacuum-metal interfaces. 
The electric field is evaluated self-consistently including the correct screening of the free charges at the inner surface. As a result, the screening term $\frac{2}{1+ \epsilon_r} $, which anyway is a crude approximation here, is no longer necessary. 

Second, the charge relaxation is controlled by the bulk and   surface conductivity of the insulator, which may even depend non-linearly on  the electric field. The charges that migrate from the inner to the outer surface accumulate at the outer surface and still contribute to the electric field. Absorbing boundary conditions at the outer surface can be added to permit the accumulated charges to decay, simulating a leakage current that allows to reproduce the measured charge decay time. The grid method allows thus to describe quite reliably the charge dynamics in insulating capillaries, which is crucial in order to get accurate beam particle trajectories and consequently accurate charge deposition. Reversely, as the transmitted particles monitor the time evolution of the electric field inside the capillary, one has also access to  some aspects of the charge dynamics in insulators.  If the comparison between experimental results  and simulations are reliable enough, capillaries may thus also set to become a formidable tool to study the charge dynamics in insulators.

Last but not least, we consider the CPU cost for the charge dynamics. In the point charge approach, when the $N$ point charges are field driven, the electric field needs to be evaluated at the position of each point charge, requiring thus $N^2$ evaluations. With increasing number of accumulated point charges, the charge dynamics  may quickly became a bottleneck in the simulation. In the grid approach, the CPU cost for the charge dynamics is constant as the charge dynamics does not depend on the number of accumulated charges in the insulator but only on the number of grid points. The grid approach is thus well suited for studies where the insulator accumulates  a large number of charges.   

The grid  approach was used for example to simulate the self-organized radial focusing of a 2.5 keV ion beam by a conical capillary \cite{Giglio_PRA_2018,Giglio_PRA_emittence}. Further it  was used to show in how far  secondary electrons, generated by ions hitting the capillary inner surface, can avoid Coulomb blocking of the transmitted beam \cite{Dubois_PRA_2019,Giglio_PRA_2021}.     
In both cases it was necessary to simulate a large number of trajectories (about $10^7$) and accumulate a large amount of charges in the capillary to study the time evolution of the transmitted beam.

While the surface grid method has many advantages, it may not be implemented as straightforwardly as the point charge method, which may explain why the surface charge dynamics on an axial 2D grid was not widely adopted by others. However, in the special case where the capillary is a straight cylinder  and where the bulk and surface conductivities can be considered constant, one can obtain an implementation of the charge dynamics which is robust, simple and fast. In addition,  making intense use of Fast Fourier Transform, one obtains also a fast numerical scheme for evaluating the electric field at a beam particle's position, which is relevant for the computation of the trajectory of the beam particles. 

In this work we will present such an implementation, based on the spectral method. Instead of using a cylindrical 2D grid to describe the charge dynamics, all the calculations are done using the moments of the distributions. We emphasize that the numerical schemes proposed in this work are suited only for straight capillaries. For example for conical capillaries, a grid method for the surface charges is still preferable.

Surface charge dynamics on a grid has not been adopted by other authors in the field and one reason may be algorithmic/numerical complexity of such an approach. This work  presents the numerical schemes that we use for simulating the guiding of ions through insulating straight nano or macro capillaries. The paper may be regarded as a walk through our numerical code where we will detail how we simulate the  inserted beam and control its emittance, how we compute the charge deposition and charge dynamics in the insulating capillary, how we evaluate the electric field by accounting for imposed boundary conditions, and how we evaluate efficiently the trajectories of the inserted charged particles.

%

\section{Theoretical Model}
The theoretical model, on which the charge dynamics is based, have already been  presented elsewhere \cite{Giglio_PRA_rates}.
In this section we will merely present  those parts of the model that are relevant for understanding the numerical schemes of the code.
 
\subsection{Parameters of the system}

\subsubsection{Dimensions of the capillary}
\begin{figure}[ht!]
\begin{center}
\includegraphics[scale=0.5]{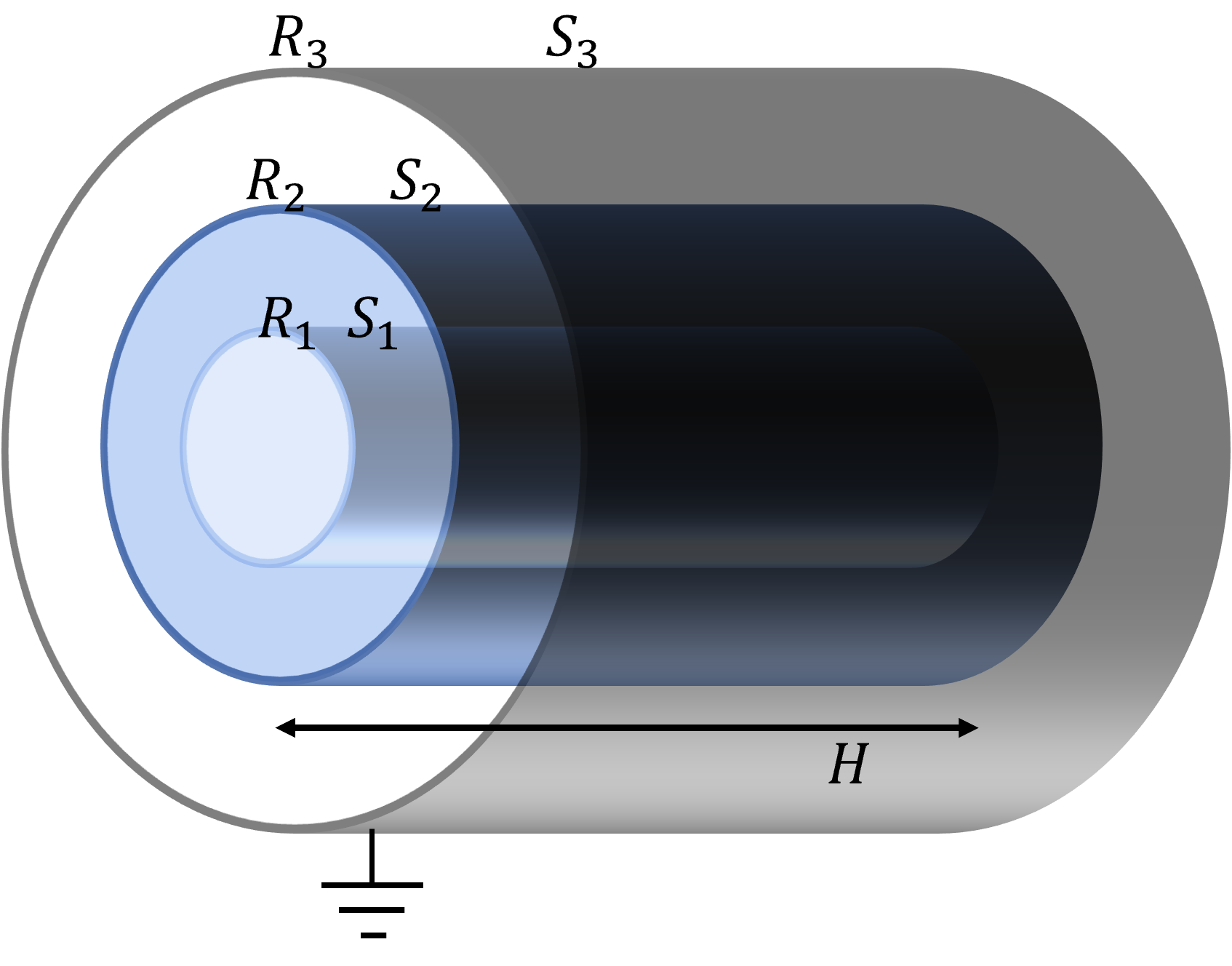}
\end{center}
\caption{The three interfaces $S_1$, $S_2$, and $S_3$ of the capillary setup. The conducting $S_3$ interface is grounded.}
\label{fig_scheme}
\end{figure}

A straight insulating capillary has a length $H$, an inner radius $R_1$ and an outer radius $R_2$, as depicted in Fig.\ref{fig_scheme}. We label $S_1$ the inner vacuum-dielectric interface and $S_2$ the outer dielectric-vacuum interface. We further suppose that the capillary is surrounded by a metallic cylindrical surface $S_3$  of  radius $R_3\ge R_2$ and length $H$.  The interface $S_3$ shares the same symmetry axis than the capillary and is electrically grounded. Surface $S_3$ is an important addition because it imposes Dirichlet boundary conditions on  the electric potential outside the capillary, which allows  determining the electric potential unambiguously. In experiments, a grounded metal cylinder should therefore always surround the capillary, if a reliable comparison between experiments and theory is desired. 

In the particular case where the outer surface ($S_2$) of the capillary is covered by a conducting paint and grounded like in \cite{Gruber12,Gruber2014}, it is sufficient to put $R_3=R_2$.  On contrary, if no grounded cylinder is  present in the experiment, we recommend to use $R_3=H$, which should be large enough so that $S_3$ does not influence significantly the electric field inside the capillary. 
In experiments, the front circular section of the capillary is grounded or mounted behind a grounded collimator plate with an aperture centered on the capillary entrance. 

\label{sec_theo}

\subsubsection{Electric properties of the capillary}

The model assumes that the deposited charges do not accumulate in the bulk but are localized at the inner and outer interface of the capillary. This assumption should be valid for insulators where the electric conductivity is dominated by ionic conductivity, such as glasses.
Indeed, when beam particles hit the interface $S_1$, electrons are ejected from the impact point, or equivalently,  holes are injected at the impact point. Those holes are considered to have a negligible mobility compared to alkali ions and the holes are mainly relaxed by ionic conductivity. And because ionic charge carriers cannot pass the interfaces $S_1$ and $S_2$, a positive charge builds up at the interfaces, which is  well approximated by surface charge densities at $S_1$ and $S_2$. Our model assumes that the electric bulk and surface conductivities are constant in time and space. Consequently, field dependent electric conductivities are not supported. 

The electric properties of the capillary are  described by three parameters,  the relative permittivity $\varepsilon_r$, the bulk conductivity $\kappa_b$ and the surface conductivity $\kappa_s$. 
Note that the latter is mainly due to absorbed impurities and surface defects. The author recommends therefore to use, if possible, values for the surface conductivity which were measured in secondary vacuum. But in general the surface conductivity is not a well-known quantity and its contribution to the charge relaxation rate is still an open question. Note that the surface conductivity on $S_1$ and $S_2$ are not necessarily equal and the reader may use different values for both.

\subsubsection{Symmetries and boundary conditions}
We assume here that the capillary axis is aligned with the z-axis of the cylindrical coordinate system $(r,\theta,z)$ and that the beam axis lies in the xOz plane.  The deposited charge distributions and consequently the electric potential will thus have xOz plane symmetry. 

At the interfaces $S_1$ and $S_2$ and $S_3$, boundary conditions are imposed.  At $S_1$ and $S_2$, the normal components of the electric displacement field are discontinuous, with their difference equal to the respective free surface charge densities.  At the grounded  interface $S_3$, the potential is zero. 
\begin{eqnarray}
 \varepsilon_r \varepsilon_0 E_r(R_1^+) - \varepsilon_0 E_r(R_1^-) &=& \sigma^{(1)} \label{BC_R1}\\
\varepsilon_0 E_r(R_2^+) - \varepsilon_r \varepsilon_0 E_r(R_2^-) &=& \sigma^{(2)} \label{BC_R2}\\
V(R_3) = 0\label{BC_R3}
\end{eqnarray}
Equations (\ref{BC_R1}-\ref{BC_R3}) add  surface polarization charges at both  dielectric interfaces and free charges at  the conducting interface $S_3$.
To determine the electric potential unambiguously, we also have to impose boundary conditions at $z=0$ and $z=H$. 
The front circular section of the capillary is supposedly covered by a conducting layer or a collimator plate and grounded. We put thus the electric potential at $z=0$  equal to zero. Further we assume an ohmic contact between the grounded layer and the front circular section of the capillary so that electrons can flow from the grounded layer to the front circular section of the capillary and neutralize the excess charges that gather at $z=0$. Consequently, we impose an absorbing boundary condition at $z=0$ for the  surface charge densities $\sigma^{(1)}$ and $\sigma^{(2)}$ at the inner and outer surface respectively, 
\begin{gather}
\left.
\begin{matrix}
 V &= 0  \\ 
 \sigma^{(1,2)} &= 0
\end{matrix} 
\right\}
\quad z=0
\end{gather}
At the rear circular section of the capillary ($z=H$), the reader can choose between an Dirichlet type  or Neumann type boundary  boundary condition.  
\begin{enumerate}
\item  If the rear circular section of the capillary is covered by a conducting layer and grounded, which is usually the case  for nano-capillaries, we will chose a Dirichlet type boundary condition. In that case, the electric potential $V$ and the surface charge densities $\sigma^{(1,2)}$ are set to  zero at $z=H$. Imposing the charge density to be zero at the boundary  implies that the charge densities are absorbed at the grounded rear circular section.

\item If the rear circular section of the capillary is electrically isolated, which is often the case for macro-capillaries, we may choose a Neumann boundary condition. In  that case the z-component of the electric field $\frac{\partial V}{\partial z}$ and the z-component of the surface charge current $\frac{\partial \sigma}{\partial z}$ are set to zero at $z=H$. The latter implies that the charges cannot pass the boundary (blocking boundary).

\end{enumerate}  
The Neumann condition seems to be the natural choice if the rear circular section is not explicitly grounded. But if a positively charged capillary is not screened from stray electrons then the positive potential of the rear circular section may attract stray electrons until the potential at $z=H$ is close to zero \cite{Giglio_PRA_2017}. In that case, a Dirichlet boundary condition at $z=H$ is probably better suited.

\subsection{Spectral method}
\label{sect_spectral}

Discrete Fourier transforms (DFT) expand a function $f(\vec{r})$ on orthogonal circular basis functions.
In this Fourier representation, differentiation is straightforward and a linear differential equation with constant coefficients is transformed into an easily solvable algebraic equation. The inverse DFT allows then to transform the result back into the ordinary spatial representation. Such an approach is called a spectral method.
One advantage of the spectral method is that the circular basis functions used in the expansion can be chosen to automatically satisfy imposed symmetries and boundary conditions on $f(\vec{r})$.  In the following we define the quantities that are relevant for the charge dynamics.

\subsubsection{Surface charge distributions}
In cylindrical coordinates $(r,\theta,z)$, 
the surface charge densities at the inner $\sigma^{(1)}$ and outer surface $\sigma^{(2)}$ are expanded on the following circular basis functions
\begin{eqnarray}
\sigma^{(1,2)}(\theta,z,t) & = & \sum_{m=0}^M \sum_{n=1}^N  \sigma^{(1,2)}_{mn}(t) \cos(m \theta)  \sin(k_n z) \label{expans_sig1} \quad .\notag \\
\end{eqnarray}
The orthogonal basis functions $\{ \cos(m \theta)\} $  account for the xOz plane symmetry of the problem, while the $\{ \sin(k_n z) \}$ functions satisfy automatically  the absorbing boundary condition at  the capillary inlet $z=0$.
The $\sigma_{mn}^{(1,2)}(t)$ are the time dependent moments of the distributions, where $m$ is the angular index and $k_n$ the wave number of index $n$. The definition of $k_n$ differs according to the imposed boundary condition at $z=H$. 
If we define $k_n$ by 
\begin{equation}
k_n= n \frac{ \pi}{H} \quad, \forall \, n \ge 1\quad,
\end{equation}
then $\sin(k_n H)=0$, which imposes an absorbing boundary condition at the rear circular section of the capillary.
If we define $k_n$ by 
\begin{equation}
k_n= (n-\frac{1}{2})\frac{ \pi}{H} \quad, \forall \, n \ge 1\quad,
\end{equation}
then $\partial_z \left.\sin(k_n z) \right|_{z=H} =0$, which imposes a zero-flux (blocking) boundary condition at $z=H$.
In both cases, the $\{ \sin(k_n z) \}$ functions form an orthogonal basis set on the interval $[0,H]$.
Consequently, knowing the moments $\sigma_{mn}^{(1,2)}(t)$ is equivalent to knowing the densities $\sigma^{(1,2)}(\theta,z,t)$ at the interfaces of the tube, with the addition that the imposed  boundary conditions at $z=0$ and $z=H$ are automatically satisfied.


\subsubsection{Source distributions}
When beam particles hit the surface, a charge is deposited at the impact point. We label $\gamma^{(1)}$ and $\gamma^{(2)}$  the deposited charge per unit time and unit area at the inner and outer surface, respectively. In the following we will refer to $\gamma^{(1,2)}$ as source distributions.  They are expanded similarly,
\begin{eqnarray}
\gamma^{(1,2)}(\theta,z,t) & = & \sum_{m=0}^M \sum_{n=1}^N  \gamma^{(1,2)}_{mn}(t) \cos(m \theta)  \sin(k_n z) \quad. \notag \\
\end{eqnarray}
inheriting thus also the $xOz$ plane symmetry and boundary conditions properties of the circular basis functions. 
\subsubsection{Electric potential inside the capillary}

Inside the capillary, $0 \le r\le  R_1$, there are no charges and the electric potential satisfies the  Laplace equation. The potential is thus expanded on harmonic basis functions. Retaining only those basis functions which well-behave at $r=0$, satisfy the $xOz$ plane symmetry and imposed boundary conditions at $z=0$ and $z=$, one has,
\begin{align}
V(r,\theta,z,t)= \sum_{{m=0} \atop {n=1}}^{M,N} v_{mn}(t) I_m(k_nr) &  \cos(m \theta)  \sin(k_n z) \notag   \\  \label{eq_app_v1_sigmas}
\end{align}
where $I_m()$ are modified Bessel functions of order $m$. Depending on the chosen definition of $k_n$, the potential or the z-component of the electric field are zero at $z=H$. The terms $v_{mn}(t)$
depend linearly on the time-dependent surface charge moments \cite{Giglio_PRA_rates},
\begin{equation}
v_{mn}(t) =  a_{mn} \sigma^{(1)}_{mn}(t) + a'_{mn}\sigma^{(2)}_{mn}(t) \quad . \label{vmndef}
\end{equation}
The coefficients $a_{mn}$ and $a'_{mn}$,
are  constants and depend merely on the parameters $R_1,R_2, R_3, H,\varepsilon_r$ of the system.  They  account for polarization charges that appear at the two vacuum-dielectric interfaces ($S_1$) and ($S_2$)  and at the vacuum-metal interface ($S_3$). In appendix \ref{appendixPmn}, we give a Python code that generates the $a_{mn}$ and $a'_{mn}$ coefficients. Once those coefficients given, the potential $V$ and the electric field  $-\vec{\nabla} V$  inside the capillary tube are readily computed  using (\ref{vmndef}) and (\ref{eq_app_v1_sigmas}).

Strictly speaking, imposing Dirichlet or Neumann boundary conditions on the potential $V$   at $z=0$ and $z=H$ introduces smalls errors on the electric field near the inlet and outlet of the capillary. For a Dirichlet boundary condition at $z=0$ or   $z=H$, the error can be corrected by adding to the potential $V()$ the  potential generated by a disk of radius $R_1$, placed at the boundary and carrying the surface charge distribution $\varepsilon_0  \frac{\partial V}{\partial z} $. For the Neumann boundary condition at $z=H$,  the error can be corrected by adding to the potential $V$ the electric potential generated by a disk of radius $R_1$, placed at the boundary and carrying the potential $-V(r,\theta,z=H)/2 $. Fortunately,  the error is limited to the regions near the inlet and outlet of the capillary and, because of the large aspect ratio of the capillary, it does not affect sensibly the trajectories of the beam particles and can thus usually be ignored.
%
%

\subsection{Time evolution of the surface charge moments}

In the case where the charge dynamics takes place on the interfaces of a straight cylindrical capillary and where the surface and bulk conductivities are constants, the  charge dynamics can actually be mapped onto a set of linear differential equations \cite{Giglio_PRA_rates},
\begin{gather}
\begin{pmatrix}
\dot{\sigma}^{(1)}_{mn}(t) \\ \\ \dot{\sigma}^{(2)}_{mn}(t)\\ 
\end{pmatrix} 
= -
\underbrace{\begin{pmatrix}
f^{({11})}_{mn} \quad f^{({12})}_{mn}  \\ \\
f^{({21})}_{mn} \quad f^{({22})}_{mn} 
\end{pmatrix} }_{F_{mn}}
\underbrace{
\begin{pmatrix}
\sigma^{(1)}_{mn} (t)\\ \\ \sigma^{(2)}_{mn}(t)\\
\end{pmatrix} }_{\vec{\sigma}_{mn}(t)}
+
\underbrace{
\begin{pmatrix}
\gamma^{(1)}_{mn}(t) \\ \\ \gamma^{(2)}_{mn}(t)\\
\end{pmatrix} }_{\vec{\gamma}_{mn}(t)} \label{eq_System}
\end{gather}
The time-{\bf in}dependent coefficients  $f^{({ij})}_{mn}$ are rates
and depend on the electric properties and dimensions of the capillary, $R_1, R_2, H, \kappa_b, \kappa_s$ and $\varepsilon_r$. The rates  account for charge relaxation due to conductivity as well as for charge decay due to absorbing boundaries. Their expressions are given in \cite{Giglio_PRA_rates,Giglio_PRA_2021}. 
Remarkably, there is no coupling between  the different  modes and (\ref{eq_System}) can be solved independently for each couple $(m,n)$. 
The eigenvalues of $F_{mn}$ correspond to the two charge relaxation rates that characterize the straight capillary and are labeled respectively  $1/\tau_{mn}^{(1)}$ and  $1/\tau_{mn}^{(2)}$. 
  
Let $\delta t $ be the time step in the numerical integration of the charge dynamics. We introduce the discrete time $t_i$ define by
\begin{equation}
t_i = i \times \delta t \quad.
\end{equation}  
If we approximate the source term $\gamma_{mn}(t)$ in the time interval $ t_i\le t \le t_{i+1} $ by its time-averaged value $\overline{\gamma}_{mn}(t_i)$, see  (\ref{eq_mom_dep_charg}), 
then the set of differential equations (\ref{eq_System}) can be solved analytically. 
For better readability we introduce the auxiliary quantities,
\begin{eqnarray}
\lambda^{(i)}_{mn} &=& e^{-\delta t / \tau_{mn}^{(i)}}  \\
\Lambda^{(i)}_{mn} &=& \tau_{mn}^{(i)}(1-\lambda^{(i)}_{mn}) 
\end{eqnarray}
where $ i=1,2$. Note that $\lambda_{mn}$ and thus $\Lambda_{mn}$  depend on the time step $\delta t$.  Further, we introduce the  projectors  $P_{mn}$ and $Q_{mn}$, defined in \cite{Giglio_PRA_rates}. They are deduced from $F_{mn}$ and thus  also time-independent.
The density moments at time $t_i +\delta t$ are then given by the rather simple expression

\begin{eqnarray}
\vec{\sigma}_{mn}(t_{i+1}) &= &  (\lambda^{(1)}_{mn} \, P_{mn} +\lambda^{(2)}_{mn} \, Q_{mn}) \, \vec{\sigma}_{mn}(t_i)+ \notag \\
  & &  (\Lambda^{(1)}_{mn}  \, P_{mn} + \Lambda^{(2)}_{mn} \, Q_{mn})\,\vec{\bar{\gamma}}_{mn}(t_i) \quad. \notag \label{eq_charge_dyn} \\
\end{eqnarray}
The time evolution of the density moments can thus be  computed straightforwardly using a CPU inexpensive algorithm which is also unconditionally stable. The quantities $\lambda_{mn}^{(1,2)}$, $P_{mn}$ and $Q_{mn}$ need to be computed just once for all.
Given the parameters $R_1, R_2, R_3, H, \kappa_b, \kappa_s$ and $\varepsilon_r$, as well as $N$ and $M$, the python routine \emph{rates.py} discussed in appendix \ref{appendixPmn}, calculates the matrices $F_{mn}$ for each couple $(mn)$ then diagonalizes each matrix $F_{mn}$, extracts the two corresponding relaxation rates $\tau^{(1)}_{mn}$ and $\tau^{(2)}_{mn}$, and generates $P_{mn}$ and $Q_{mn}$. With the latter given, implementing the charge dynamics is straightforward.

\subsection{Computation of the source distributions}

\subsubsection{Inner surface}

In the following, we calculate the time-averaged source terms $\bar{\gamma}^{(1,2)}_{mn}$ introduced in the previous section. 
We assume  that all beam particles have the same charge $q$ and that the inserted beam current $I_\text{in}$ is constant in time. 
When a particles hits the inner surface $S_1$ of the capillary at a grazing angle, a charge 
\begin{equation}
q_\text{dep}=  (q+N_\text{se} \,e )
\end{equation}
is deposited at the impact point $(\theta_p,z_p)$, where  $N_\text{se}$ is the average number of emitted secondary electrons per impact and $e$ the elementary charge.  We suppose  that the deposited charge $q_\text{dep}$ is  smeared  over the interface according to a Gaussian distribution with a variance of $\Delta z = H/N $ along the $z$  axis and a variance of $\Delta \theta = \pi /M$ along the angle $\theta$. The deposited point charge is thus replaced by a deposited surface charge,
\begin{equation}
 \sigma^\text{dep}_{\theta_p,z_p}(\theta,z) =  \frac{q_\text{dep}}{\pi R_1 \Delta \theta \Delta z} e^{\frac{-(\theta -\theta_p)^2}{\Delta \theta ^2}} \, e^{\frac{-(z -z_p)^2}{\Delta z ^2}} \quad.
\end{equation}
%
Let $N_\delta \ge 1$ be the number of particles that are inserted into the capillary during the time step $\delta t$, 
\begin{equation}
  N_\delta  = \delta t \, \frac{I_\text{in}}{q} \quad \text{with} \quad  \delta t \ge \frac{q}{I_\text{in}}  \quad. 
\label{def_N_delta}
\end{equation} 
Out of the $N_\delta$ inserted particles, $N_p(t_i)$ hit the inner surface during the time $t_i$ and $t_{i+1}$. The deposited surface charge  density per unit time,  averaged over the time interval $\delta t$, reads 
\begin{eqnarray}
\bar{\gamma}^h (\theta,z,t_i) &= & \frac{1}{\delta t} \sum_{p=1}^{N_p(t_i)} \sigma^\text{dep}_{\theta_p,z_p}(\theta,z) 
\label{average_depos2}
\end{eqnarray}
where the bar symbol stands for the time average.
Projecting (\ref{average_depos2}) on the  basis functions  $\{ \cos(m \theta)  \sin(k_n z) \}$ yields the time-averaged moments $\bar{\gamma}_{mn}^h (t_i)$, defined as
\begin{align}
\bar{\gamma}_{mn}^h(t_i) & =  \int_{-\infty}^\infty d\theta \int_{-\infty}^\infty \bar{\gamma}^h (\theta,z,t_i)  \cos(m \theta)  \sin(k_n z) dz  \notag \\ = & \;   g_{mn} \times \frac{1}{N_\delta}\sum_{p=1}^{N_p(t_i)} \cos(m\,  \theta_p) \sin(k_n \, z_p) \label{eq_mom_dep_charg} \quad,
\end{align}
where the factors
\begin{equation}
g_{mn}=  I_\text{in}  \left( 1 + \frac{N_\text{se} e  }{q  }  \right)  \frac{4    }{\pi R_1 H} 
e^{-\frac{1}{4}(\Delta \theta^2 m^2+ \Delta z^2 k_n^2)} 
\label{eq_gmn}
\end{equation}
are computed once for all for to save CPU time. Secondary electrons, emitted from the impact point, are field driven  and eventually absorbed elsewhere at the the interface $S_1$ \cite{Giglio_PRA_2021}. 
The source term $\bar{\gamma}_{mn}^{\text{se}}(t_i)$ accounts for the deposited secondary electrons distribution at time step $t_i$. Summing both contributions yields the total source moments at the inner surface, 
\begin{equation}
\bar{\gamma}_{mn}^{(1)}(t_i) = \bar{\gamma}_{mn}^{h}(t_i) + \bar{\gamma}_{mn}^{\text{se}}(t_i) 
\end{equation}

\subsubsection{Outer surface}
 
For the outer surface $S_2$, we distinguish tree cases. If $S_2$ is covered by the conducting layer $S_3$ and grounded, $R_2=R_3$,
no surface charges accumulate at the interface $S_2$ and we can safely put
\begin{equation}
\sigma_{mn}^{(2)}(t_i) =0 \quad \text{and } \quad \bar{\gamma}_{mn}^{(2)}(t_i)=0
\label{grounded_R2}
\end{equation}
Equation (\ref{grounded_R2}) represents thus an absorbing boundary condition at $S_2$.

If $R_3> R_2$ and if  $S_2$ is sufficiently screened from stray electrons, we can safely put  $\bar{\gamma}_{mn}^{(2)}(t_i)=0$.
Otherwise, stray electrons may be attracted by the charged capillary, see \cite{Giglio_PRA_2017}, and a stray electron current per unit surface $\bar{\gamma}^{(2)}(\theta,z,t_i)$ must be defined at $S_2$ by the reader. 
\subsubsection{Computational cost}
While (\ref{eq_charge_dyn}) is unconditionally stable, the time step $\delta t$ has nevertheless  to satisfy the following condition. 
The source terms $\gamma(t)$ should vary sufficiently slowly during the time step $\delta t$ so that is can be approached by its time-averaged value. That implies that the  electric field inside the capillary should not vary significantly during the time step, which is ensured if $\delta t$ is small compared to the characteristic relaxation times $\tau_{mn}^{(1,2)}$ and if the inserted charge per time step does not exceed a certain value $Q_\delta$, 
\begin{equation}
I_\text{in} \,\delta t = q N_\delta \le  Q_\delta \quad ,
\label{insert_Q}
\end{equation}
where the value of $Q_\delta$ depends on the dimensions of the capillary.
%
%
For nano-capillaries, we estimate that $Q_\delta$ should not  exceed several elementary charges. With inserted currents  typically in  the range of $10^{-18}$A, we take  $N_\delta=1$ and thus $\delta t = q/ I_\text{in}$.

For  macro-capillaries, we estimate that $Q_\delta$ should not  exceed  $10^{-14} C$.  With inserted currents typically in the range of $10^{-13}$A, equation (\ref{insert_Q}) yields a time step of the order of $\delta t \sim 0.1$s and about $N_\delta\sim 10^5$ inserted particles per time step $\delta t$.   This is an unnecessarily large number of trajectories that need to be computed per time step $\delta t$ to evaluate the time-averaged source term, see Eq. (\ref{eq_mom_dep_charg}). In our simulations we therefore state that
%
one calculated trajectory stands for $y\ge 1$ physical particles.
The number of trajectories $N_\delta$ that need to be calculated per time step is thus reduced by the factor $y$ 
\begin{eqnarray}
N_\delta &=& \left \lfloor \frac{Q_\delta}{y \, q} \right \rfloor 
\end{eqnarray}
while equation (\ref{def_N_delta}) is replaced by 
\begin{eqnarray}
\delta t &= & N_\delta \,\frac{ y \,q}{I_\text{in}} 
\end{eqnarray}

While larger values of $y$ reduce the number of the trajectories to be computed per time step, too large values for $y$ may introduce spurious effects due to low statistics in the evaluation of the time-averaged source terms $\bar{\gamma}^h_{mn}(t_i)$ (\ref{eq_mom_dep_charg}).  We recommend to keep $y$ below 5000.
For inserted currents  $I_\text{in}\ge 10^{-15}$A, a thumb rule for $y$ may be given by the relation
\begin{equation}
y = 10^{-3} \times \frac{I_\text{in}}{q} \quad  .
\end{equation}

The computational cost for the evaluation of the  source moments $\bar{\gamma}_{mn}^{(h)}(t_i)$  per time step scales  thus as $N_\delta\times (1+N_\text{se}) \times M \times N$, which is relatively inexpensive,  
especially that recursive relationships can be used to 
generate  efficiently the $M$ terms  $\{\cos(m \,\theta_p)\}$ and $N$ terms $\{\sin(k_n z_p)\}$ that appear in (\ref{eq_mom_dep_charg}), see Appendix \ref{appendix_recursive}. 
As a result, the computation of the moments $\bar{\gamma}_{mn}^{(1,2)}(t_i)$ is fast with the advantage of being simple, without the need of an additional grid mapping the interface $S_1$ and $S_2$.

 \subsection{Computation of the electric field on a 3 D grid}

\subsubsection{Fast Fourier related transforms}
Unlike  charge densities and source terms, the electric field and electric potential will be be evaluated on a 3D cylindrical grid ($r_i,\theta_j,z_k$) mapping  the interior, $r_i \le R_1$, of the capillary. Along the angular and axial coordinates the grid point $\theta_j$ and $z_k$  are  uniformly distributed, 
\begin{eqnarray}
\theta_j &=& j \frac{2 \pi}{M}  \quad, \quad j \in [0,M-1] \\
z_k &=& k \frac{H}{N}  \quad,  \quad k \in [0,N] \quad .
\end{eqnarray}
This is motivated by the desire to use Fast Fourier related transforms to pass from the representation in the moment space to the grid points representation. Therefore $N$ and $M$ should be powers of 2.
As an indicator, for capillaries with an aspect ratio $H/R_1$ of less the 50, the author  uses $N=256$ and $M=16$ in his simulations. For larger aspect ratios, $N$ may be doubled.

Along the radial coordinate, no Fast Fourier related transform will be used and the radial grid is spaced non-uniformly, becoming denser with increasing  $r$. This is motivated by the fact that the electric field varies faster when approaching the surface.
Let $L+1$ be the number of radial grid points. We propose  to distribute the radial grid point according to the relation
\begin{eqnarray}
r_i &=& R_1 \sqrt{\frac{i}{L}} \quad, \quad  i \in [0,L] 
\end{eqnarray}
The grid has thus $\Omega=(L+1)\times M \times N$ grid points. The author typically uses for the radial grid $L=7$, resulting in $\Omega \sim 32000$.

The potential and electric field on the $(r_i,\theta_j,z_k)$ grid points are computed by applying the following scheme.
In a  first step we define for a given  time $t$, the auxiliary moments
$\psi_{mn}(r_i)$ and its radial derivative $\psi'_{mn}(r_i)$
\begin{eqnarray}
\psi_{mn}(r_i) &=&  v_{mn}(t) I_m(k_n r_i)  \\
\psi'_{mn}(r_i) &=& -v_{mn}(t) \left. \frac{I_m(k_n r)}{\partial r}\right|_{r=r_i}  \quad,
\end{eqnarray}
where the $v_{mn}(t)$ are given by (\ref{vmndef}).
In a second step, we pass from the moments of index $n$ to the real space points $z_k$, by applying a discrete sine (DST) or discrete cosine (DCT) transforms to the moments $\psi$ and $\psi'$ as indicated below.  
\begin{eqnarray}
\psi_{mn}(r_i)  &\xrightarrow{\text{DST}} & \xi_m(r_i,z_k) \quad \forall\; i,m \notag \\
\psi'_{mn}(r_i)  &\xrightarrow{\text{DST}} & \xi'_m(r_i,z_k) \quad \forall\; i,m \notag \\
k_n \times \psi_{mn}(r_i) & \xrightarrow{\text{DCT}} & \chi_m(r_i,z_k) \quad \forall\; i,m \notag
\end{eqnarray}
and store  the results in auxiliary quantities $\xi,\xi'$ and $\chi$.
In a third step, we pass from the angular moments $(m)$ to the angular grid points $(\theta_j)$,
\begin{eqnarray}
\xi_m(r_i,z_k) &\xrightarrow{\text{DCT}} &V(r_i,\theta_j,z_k)\quad \forall\; i,n \\
\xi'_m(r_i,z_k) &\xrightarrow{\text{DCT}} &E_r(r_i,\theta_j,z_k)\quad \forall\; i,n \\
m\times  \xi_m(r_i,z_k) &\xrightarrow{\text{DST}} &E_\theta(r_i,\theta_j,z_k)\quad \forall\; i,n \\
-\chi_m(r_i,z_k) &\xrightarrow{\text{DCT}} &E_z(r_i,\theta_j,z_k)\quad \forall\; i,n  
\end{eqnarray}
Thank to the efficiency of the DST and DCT, the CPU cost for evaluating the potential $V$ and the 3 component of the electric field $(E_r,E_\theta,E_z)$ on the 3D cylindrical grid scales like \begin{equation}
7 \times \Omega (\log(M)+\log(N))  \quad.
\end{equation}
Updating the electric field on the grid is relatively CPU expensive  but  fortunately  it  does not need to be carried out  every time step. Only when a relative change in the  moments $v_{mn}$ exceeds a certain amount  $ \epsilon >0$ does one need to re-evaluated the electric field at the grid points. 
Let $t_j$ be the last time when the electric field has been evaluated on the grid. If at a later  $t_i>t_j$ there is at least one couple of indexes ($m,n$) for which
\begin{equation}
\left| v_{mn}(t_i) -v_{mn}(t_j)  \right| > \epsilon \left| v_{mn}(t_j) \right|
\end{equation}
then and only then will the electric field on the grid be updated with the new values taken at time $t_i$. As a result, the frequency with which the field is updated on the grid decreases with increasing amount of accumulated charge in the capillary. The value of  parameter $\epsilon$ is fixed by the reader. The author typically  uses $\epsilon = 1\%$.

\section{Trajectories}

\subsection{Equation of motion}

Particles of  charge $q$ and mass $m$, extracted by a potential $V_s$ have an initial velocity of $u_0=\sqrt{2 q V_s/m}$. Here we assume that only the electric field generated by the deposited charge in the capillary wall acts on the charged particles. Other forces like the image charge force at the vacuum-dielectric interface may be added by the reader \cite{Giglio_image_force}. The trajectories of charged beam particles are calculated using classical Newtonian dynamics, which in its dimensionless form reads \cite{Giglio_PRA_2021}
\begin{equation}
\begin{matrix}
\frac{d \vec{\tilde{u}}}{ d \tilde {t}} = -\frac{1}{2}  \frac{\partial \tilde{V}}{\partial \tilde{\vec{r}}} \\ \\
\frac{d \vec{\tilde{r}}}{ d \tilde {t}}= \vec{\tilde{u}}
\end{matrix} 
 \quad,\quad  \text{with} \;\left|\begin{matrix}
\vec{\tilde{u}}=&\vec{u}/u_0\\
\vec{\tilde{r}}=&\vec{r}/H\\
\tilde{t}=&t \, (u_0/H) \\
\tilde{V} = & (V/V_s)
\end{matrix}
\right.
\label{eq_dimless_EOM}
\end{equation}
where the tilded quantities are dimensionless. 
Note that the equation of motion (EOM) is independent from the charge over mass ratio $q/m$ of the beam particles. The trajectory of a beam particle is thus also independent from  $q/m$ and depends only on its initial position $(x_0,y_0,0)$ and  the initial velocity vector $\vec{u}$.
The initial conditions are  sampled by a rejection method which will be discussed in the next section \ref{emittance}.
Indeed, some care must taken when choosing the initial position and  velocity vector of the beam particles, as the emittance of the injected beam was found  to have in some cases a strong influence on the transmission rate \cite{Giglio_PRA_emittence}. The initial  conditions should therefore be sampled such as  to reproduce at best the emittance of inserted beam in the experiments. 

\subsection{Initial phase-space conditions}
\label{emittance}

The parameters of the beam that enter the simulation are: 
\begin{itemize}[parsep=0cm]
\item the injected intensity $I_{\text{in}}$,
\item the extraction potential $V_s$ of the source, 
\item the mass $m$ the particles, 
\item the charge $q$ of the particles,
\item the tilt angle  $ \beta $ between the capillary axis and the beam axis
\item the divergence (opening angle) $\alpha$ of the beam,
\item the radius $R_s$ of the virtual source and distance $D$ of the virtual source from the capillary entrance. 
\end{itemize}
With the above parameters, the sampling of the initial positions and velocities of the inserted beam particles is fully controlled.

We remind that the entrance of the capillary is placed at $z=0$. The components $(x_0,y_0,0)$ of the initial position and  $(u_x,u_y,u_z)$ of the initial velocity vector are sampled  by a rejection method according to the  following scheme, already discussed in \cite{Giglio_PRA_emittence}.
Let us assume that the profile of the beam in the plane $z=0$  has a Gaussian shape, centered on the capillary aperture and  with a full width half maximum (FWHM) labeled $W$.
The charged particles are emitted  from a virtual source, located at a distance $D$ upstream from the capillary entrance. The ratio $W/D =\tan \alpha \simeq \alpha \ll 1$ may be regarded as the divergence of the beam.

The virtual source, which   corresponds a the focal waist of the beam, is modeled by a disc of radius  $w_s$. We typically use $w_s=1$ mm. The positions ($x_s,y_s$) of the particles on the virtual source disc are sampled uniformly. The $u_x$ and $u_y$ components are sampled according to a normal distribution $P()$ having a width of $\Delta u=\alpha \, u_0 $,
\begin{eqnarray}
P(u_i) &=& \frac{\sqrt{\pi}}{\Delta u}e^{ -u_i^2 /\Delta u^2 } \quad, \quad i=x,y\\
u_z &=& u_0
\end{eqnarray}
Assuming that no electromagnetic field is present between the virtual source and the cappilary's entrance, the particles travel in a straight line until they reach  the plane $z=0$, defining initial coordinates 
\begin{eqnarray}
x_0 &=& x_s+u_x\frac{d}{u_0} \\
y_0 &=& y_s+u_y\frac{d}{u_0} \\
z_0 &=& 0
\end{eqnarray}
Only the trajectories that pass through the  aperture, $x_0^2+y_0^2<R_1^2$, are retained and constitute the initial conditions for  the equation of motion (\ref{eq_dimless_EOM}). 

In the present model, the emittence of the beam that enters the capillary is thus controlled by three parameters,
namely the distance $D$ of the virtual source (focal waist) from the aperture, the radius  $w_s$ of the virtual source disk and the divergence of the beam $\alpha$, which are easily accessible in experiments.
Eventually, the EOM is integrated numerically  using  
a $4^{th}$ order explicit Beeman algorithm with adaptive step-size control, which was designed to allow for a high number of particles in  molecular dynamics simulations  \cite{Beeman}.

\subsection{Interpolation of the electric field at the particle's position}

In order to compute  the trajectory of a beam particle trough the capillary, one needs to evaluate the electric field at a large number of different positions. As the electric field is given at the $\Omega$ grid points of a 3D  cylindrical grid mapping the inner space of a straight capillary, the electric field at given position can be obtained simply by a 3D interpolation. 
The simplest approach is the trilinear interpolation, where the 8 nearest points from a given position $\vec{r}$ are used.  Trilinear interpolation needs 7 calls to a one-linear interpolation function.
We recommend however a tricubic interpolation scheme where the 64 nearest points are used. It needs 21 calls to a one-dimensional cubic interpolation function but will yield much better energy conservation along the trajectory than a trilinear scheme.

\section{Conclusion}

We present a walk through our numerical code InCa4D where we show how to implement the surface charge dynamics at the inner and outer surface of a straight insulating capillary, without the use of cylindrical 2D grids.  The advantages of such an approach with respect to point charges  are highlighted. In particular, by satisfying the imposed boundary conditions, the electric field in InCa4D is evaluated accurately, yielding the correct charge relaxation and charge depletion rates of the deposited charges. The surface dynamics approach  is suited for both macro and nano capillary but its CPU efficiency is mostly appreciable for macro capillaries where the number of deposited charges may be large.  

We also show how to sample the initial positions and velocities of the beam particles as a function of the diameter of the virtual source disk, its distance to the capillary entrance and the divergence of the beam.  Finally we show how to compute efficiently the trajectories of the beam particles through the capillary. The key idea is to map the inner space of the capillary by a cylindrical 3D grid, then use Fast Fourier related transforms to evaluate the electric field at the grid points. Then the field at the position of the particles is obtained  using, for example, a tricubic interpolation scheme.  
This work hopefully shows that it is rather easy and advantageous to implement the  charge dynamics at the inner and outer surface of a straight capillary using the moments of the distributions. 
In a future step, will extend the charge dynamics to the bulk, so the the charges can also accumulate in the bulk and not only at the interfaces, which should be relevant for those cases where the hole mobility cannot be neglected.

\section{Apendix}
%
%
%
%
%
%

\subsection{Numerical routine rates.py}
\label{appendixPmn}

The Python3 routine \emph{rates.py} calculates all the constant coefficients that are necessary to (i) simulate  via equation (\ref{eq_charge_dyn}) the charge dynamics in a straight insulating capillary and (ii) evaluate the electric potential inside the capillary using (\ref{eq_app_v1_sigmas}).
The code is freely available on \href{https://github.com/giglioeric/Capillary}{github.com}. 
Given the inputs $R_1,R_2,R_3,H,\varepsilon_r,\kappa_b, \kappa_s,M,N$ and flag $f_{bc}$,  the routine \emph{rates.py} calculates for all indexes $0\le  m < M$ and $1\le n\le N$ 
\begin{itemize}

\item the coefficients $a_{mn}$ and $a'_{mn}$ that appear in Eq. (\ref{vmndef}) and that allow to express the electric potential inside de capillary as a function of the moments of the charge distributions. 

\item the two characteristic relaxation times  $\tau_{mn}^{(1)}$  and $\tau_{mn}^{(2)}$ of the system.

\item the projector $P_{mn}$
\end{itemize}
The flag $f_{bc}$ indicates if blocking or absorbing boundary conditions are used at $z=H$ . The projector $Q_{mn}$  is deduced from $P_{mn}$ via the relation 
\begin{equation}
Q_{mn} = I -P_{mn}
\end{equation}
where $I$ is the  $2 \times 2$ identity matrix.
The routine \emph{rates.py} is based on the definitions found in \cite{Giglio_PRA_rates}. It creates an output file 'coefficients.txt'
that contains the data ordered as follow:  

\begin{gather}
\text{------------- (m,n) ------------} \\
\begin{matrix}
a_{mn} \quad \tau^{(1)}_{mn} \quad P^{({11})}_{mn} \quad P^{({12})}_{mn}  \\ \\
a'_{mn} \quad \tau^{(2)}_{mn} \quad P^{({21})}_{mn} \quad P^{({22})}_{mn} 
\end{matrix} 
\end{gather} 
%

\subsection{Evaluating CPU-efficiently the terms $\cos(m \theta_p)$ and $\sin(k_n z_p)$}
\label{appendix_recursive}

To generate efficiently the $M$ terms  $\{\cos(m \,\theta_p)\}$ that appear in (\ref{eq_mom_dep_charg}),  
the recursive relation
\begin{equation}
\cos((m+1)\,\theta_p) = 2 \cos(\theta_p) \cos(m \theta_p)-\cos((m-1)\,\theta_p)
\label{cos_rec}
\end{equation} 
should be used, starting from  $\cos(\theta_p)$. 
For the computation of the $N$ terms $\{\sin(k_n z_p)\}$, we must distinguish between two cases: 
If $k_n= n \pi /N $  then 
the recursive relation 
\begin{equation}
\sin((n+1) \xi ) = 2 \cos(\xi)\sin( n \xi)-\sin((n-1) \xi) 
\label{sin_rec}
\end{equation}
with $\xi = \frac{\pi}{N}z_p$ should be used.
If $k_n= (n-\frac{1}{2}) \pi /N $, we  first note that
\begin{equation}
\sin( (n-\frac{1}{2}) \xi ) = -\cos(n \xi) \sin(\xi/2) + \sin(n \xi) \cos(\xi/2) \quad.
\end{equation}
We can now use the Moivre formula 
\begin{equation}
\cos (n\xi) + i \sin(n \xi) = (\cos (\xi) + i \sin( \xi))^n
\end{equation}
to generate  all the $N$  terms $\{\cos(n\xi) \}$ and $\{\sin(n\xi)\}$ recursively, starting from $\sin(\xi)$ and $\cos(\xi$).

%
%

\begin{acknowledgements}
The author would like to thank for financial support by the the French National Research
Agency (ANR) in the P2N 2012 program (ANR-12-NANO-
0008) for the PELIICAEN project.  
\end{acknowledgements}

\end{document}